# A Tapered Chalcogenide Microstructured Optical Fiber for Mid-IR Parabolic Pulse Generation: Design and Performance Study

Ajanta Barh, S. N. Ghosh, Ravi K. Varshney, and Bishnu P. Pal

*Abstract*—This paper presents a theoretical design of chalcogenide glass based tapered microstructured optical fiber (MOF) to generate high power parabolic pulses (PPs) at the mid-IR wavelength (~ 2 μm). We optimize fiber cross-section by the multipole method and studied pulse evolution by well known Symmetrized Split-Step Fourier Method. Our numerical investigation reveals the possibility of highly efficient PP generation within a very short length (~ 19 cm) of this MOF for a Gaussian input pulse of 60 W peak power and FWHM of 3.5 ps. We examined quality of the generated PP by calculating the misfit parameter including the third order dispersion and fiber loss. Further, the effects of variations in input pulse power, pulse width and pulse energy on generated PP were also studied.

*Index Terms*—Microstructured optical fiber, Fiber nonlinear optics, Ultrafast optics, Temporal parabolic pulses.

## I. INTRODUCTION

INCREASING demands for high power optical pulses for fiber-based applications have revealed optical pulse distortion and optical wave breaking with consequential growth of oscillations at the pulse's tail due to stronger nonlinear (NL) effect when operated at normal group velocity dispersion (GVD) regime [1]–[3]. However there is an interesting class of breaking-free pulses, whose temporal intensity profile is parabolic in shape and posses perfectly linear chirp across it [4]. This linear chirp helps it to maintain the rate of change of time separation same for all parts of the pulse during its propagation and hence it preserves its shape. Such parabolic pulses (PPs) can be generated in a loss-less, high gain and low dispersive media as an asymptotic solution of nonlinear Schrodinger equation (NLSE) [4], [5].

Generation of stable PPs has been the focus of much research activities over past years due to their potential applications in high power lasers and amplifiers [6], [7], Supercontinuum generations [8], and all-optical signal processing and regeneration [9].

Generations of active medium-based PPs have been demonstrated recently [5], [10]. However, the active medium-based pure PP generation is affected by amplifier spontaneous emission noise and its limited band-width. Then passive medium-based alternative approaches are found for PP generation. Most efficient approach involves use of dispersion decreasing fiber (DDF) with normal dispersion [11], [12]. Other approaches use fiber grating [13], comb-like dispersion decreasing fiber [14], solid core photonic band-gap Bragg fiber [15], fiber laser cavity [16] etc. All these proposals were applicable for today's communicating wavelength (1.55 μm) and requires long fiber lengths (~ meter to km). Recently silicon waveguide based PP generation has also reported for integrated photonic application [17].

Recently mid-IR wavelength-based (2 ~ 10 μm) photonics technology have attracted a lot of research investment due to its potential applications in defense, medical surgery, molecular "finger-print", weather forecasting, sensing, and many more [18], [19]. Thus mid-IR based high power PPs should be very important for various exciting applications in the near future.

In this paper we propose a realistic design of a dispersion decreasing MOF for efficiently producing PPs at mid-IR wavelength assuming commercially available $Tm^{3+}$-doped pulsed fiber laser [19] of 2.04 μm wavelength as input. In particular, ~ 2 μm wavelength regime is very effective for accurate cutting, welding, removing and coagulating of soft & hard biological tissues. Additionally it is also very useful for remote sensing and air monitoring and is eye-safe. Chalcogenide glass (S-Se-Te) is chosen in our study as the fiber material due to their high transparency in mid-IR wavelengths and extraordinary linear and NL properties [20]. On the other hand, their chemical durability, glass transition temperature, strength, stability etc. can be improved by doping with As, Ge, Sb, Ga for drawing an optical fiber. At present their fabrication technology is well matured though

This work relates to Department of the Navy Grant N62909-10-1-7141 issued by Office of Naval Research Global. The United States Government has royalty-free license throughout the world in all copyrightable material contained herein. A. Barh gratefully acknowledges the award of a Senior Ph.D. fellowship by CSIR (India). S. Ghosh acknowledges the financial support by DST (India) as an INSPIRE Faculty Fellow [IFA-12; PH-13].

Authors Ajanta Barh, Ravi K. Varshney, and Bishnu P. Pal are with the Department of Physics, Indian Institute of Technology Delhi, Hauz Khas, New Delhi 110016 India (e-mail: ajanta.barh@gmail.com, somiit@rediffmail.com, varshney_rk_iitd@yahoo.com, and bppal@physics.iitd.ernet.in, respectively).

S. Ghosh is with the Institute of Radio Physics and Electronics, University of Calcutta, Kolkata 700009 India; Currently on leave of absence at School of Physical and Mathematical Science, Nanyang Technological University, Singapore expensive [21]-[24]. An important feature in our methodology for realizing PP is to induce suitable decrease in dispersion along length of the fiber taper by suitably choosing the taper ratio. This is achieved by introducing an up-taper because dispersion decreases with increase in core diameter. Nonlinearity too reduces accordingly due to increasing mode effective area along length of the up-tapered fiber. Accordingly light should be launched into it from its narrower side.

With the designed dispersion and nonlinearity-tailored MOF, we have theoretically investigated feasibility of transforming a Gaussian input pulse of full width at half maximum (FWHM) of 3.5 ps having a peak power of 60W at 2.04 μm wavelength to a parabolic pulse after propagation of just ~ 19 cm length of our designed MOF.

## II. NUMERICAL MODELING

### A. Proposed Methodology of PP Generation

In an ideal loss-less optical fiber with normal GVD and hyperbolic dispersion decreasing profile, the asymptotic solution of NLSE yields a parabolic intensity profile [11], [25]. Under this condition, the propagation of optical pulses is governed by the NLSE of the form [11]

$$i\frac{\partial A}{\partial z} - \frac{\beta_2 D(z)}{2}\frac{\partial^2 A}{\partial T^2} + \gamma(z)|A|^2 A = 0 \quad (1)$$

where $A(z, T)$ is slowly varying envelop of the propagating pulse in co-moving frame, $D(z)$ is length dependent dispersion profile along the tapered length, $\beta_2$ (the 2$^{nd}$ order GVD parameter) > 0, $\gamma(z)$ is longitudinally varying NL coefficient. By making use of the co-ordinate transformation, $\xi = \int_0^z D(z')dz'$, (1) transforms to

$$i\frac{\partial A}{\partial \xi} - \frac{\beta_2}{2}\frac{\partial^2 A}{\partial T^2} + \gamma(z)|A|^2 A = i\frac{\Gamma(\xi)}{2}A \quad (2)$$

where $\Gamma(\xi) = -\frac{1}{D}\frac{dD}{d\xi} = -\frac{1}{D^2}\frac{dD}{dz}$ (3)

As $D(z)$ is a decreasing function of $z$, $\Gamma$ in (3) is > 0 and hence it functions as a gain term in (1). Thus the varrying dispersion term in the so proposed DDF functionally becomes equivalent to a varrying gain fiber amplifier with normal GVD and hence the solution of (2) becomes analogous to PP described in [5].

In the next step, we studied the effect of 3$^{rd}$ order dispersion $\beta_3$ (TOD) and fiber loss ($\alpha$) on the generated PP as these propagation factors are main obstacles for generation of pure PP. Unlike a fiber amplifier, in a DDF the TOD ($\beta_3$) and total loss ($\alpha$) grows exponentially with distance. Thus after a certain length of propagation, this DDF becomes effectively a lossy and dispersive medium instead of functioning as an effective gain medium even if $\beta_3$ and $\alpha$ are very small. Approximately this tapered fiber length ($L$) should be less than a critical length (~ $1/\alpha$) [26]. Beyond this length pulse experiences the usual linear dispersive broadening.

### B. Proposed MOF Structure

For PP generation at ~ 2 μm wavelength we choose arsenic sulphide ($As_2S_3$) based MOF geometry with a solid core surrounded by a holey cladding, consisting of 4 rings of hexagonally arranged holes embedded in $As_2S_3$ matrix. $As_2S_3$ possesses lowest transmission loss ($\alpha_T$ ~ 0.4 dB/m at 2 μm) among chalcogenide glasses and very high nonlinearity ($n_2$ ~ 4.2 x 10$^{-18}$ m$^2$/W at 2 μm) [18]. The diameter of air hole and separation between two consecutive air holes are denoted by $d$ and pitch ($\Lambda$), respectively.

First we optimize the MOF cross-section at the input end of the taper in terms of confinement loss ($\alpha_c$), $D$, $\beta_3$, and mode effective area ($A_{eff}$). $A_{eff}$ is calculated from the field profile [1], GVD parameters are calculated from real part of mode effective index and $\alpha_c$ is calculated from imaginary part of the same. To maintain low $\alpha_c$, air-filling fraction ($d/\Lambda$) and $d$ should be as large as possible. However, single-mode operation of a holey fiber requires $d/\Lambda$ should be ≤ 0.45 [27]. On the other hand, to minimize the TOD effect, dispersion slope should be as low as possible at the operating wavelength (2 μm), which also requires lower $d$ and $\Lambda$ value for $As_2S_3$ based MOF. On the other hand, low $d$ and $\Lambda$ impose greater difficulty in fabrication. Thus there is a tradeoff involved between TOD, loss and suitability of fabrication. The required dispersion decreasing profile is achieved by choosing a suitable tapered ratio for a $L$. In practice a tapered fiber can be realized during its drawing from a preform through a suitable control of the fiber draw speed [28] or through selective heating and pulling [29], [30].

Modal analysis of the proposed structure shown in Fig. 1 is carried out by using Multipole Method based CUDOS® software and the evolution of input pulse inside the tapered MOF is modeled by solving NLSE through the symmetrized split-step Fourier method [1] in MATLAB®. For the input end of our optimized linearly tapered MOF, structural parameters were $d_0 = 0.52$ μm and $\Lambda_0 = 1.3$ μm. We optimized the up-tapered ratio of MOF such that at the end of 1 m the $d_0$ transforms to $d_1$ (= 0.546 μm) and $\Lambda_0$ to $\Lambda_1$ (=1.365 μm), both of which were 1.05 times their initial values (cf. Fig. 1).

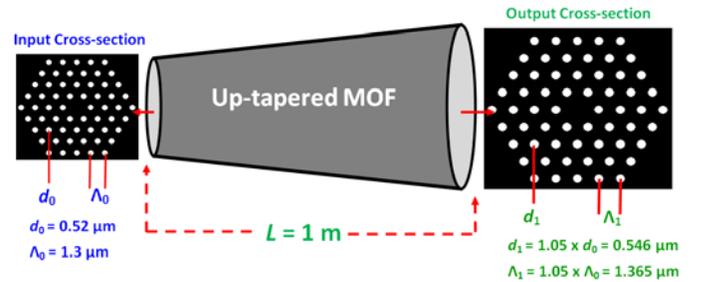

Fig. 1. Optimized linearly up-tapered MOF structure. The fiber length ($L$) considered is 1 m. At the input, the value of $d$ ($d_0$) and $\Lambda$ ($\Lambda_0$) is 0.52 μm and 1.3 μm, respectively. At the output, the value of $d$ ($d_1$) and $\Lambda$ ($\Lambda_1$) is 1.05 times their initial values (i.e. $d_1 = 0.546$ μm and $\Lambda_1 = 1.365$ μm); white circles in the figure correspond to air holes embedded in uniform $As_2S_3$ matrix (black background).

## III. RESULTS WITHOUT TOD AND LOSS

### A. Variation of Dispersion and Nonlinearity

As expected, increase in core diameter along *L* leads to a corresponding decrease in dispersion (cf. Fig. 2(a)), where as it leads to larger $A_{eff}$ and hence nonlinearity decreases (cf. Fig. 2(b)) along *L*. Value of $|D|$ is ≈ 91.78 ps/km.nm at the input end, which reduces to 82.77 ps/km.nm at the output end of the taper. On the other hand, the value of *γ* reduced from ≈ 3.69 – 3.49 (W.m)$^{-1}$ from the input to the output end over *L*.

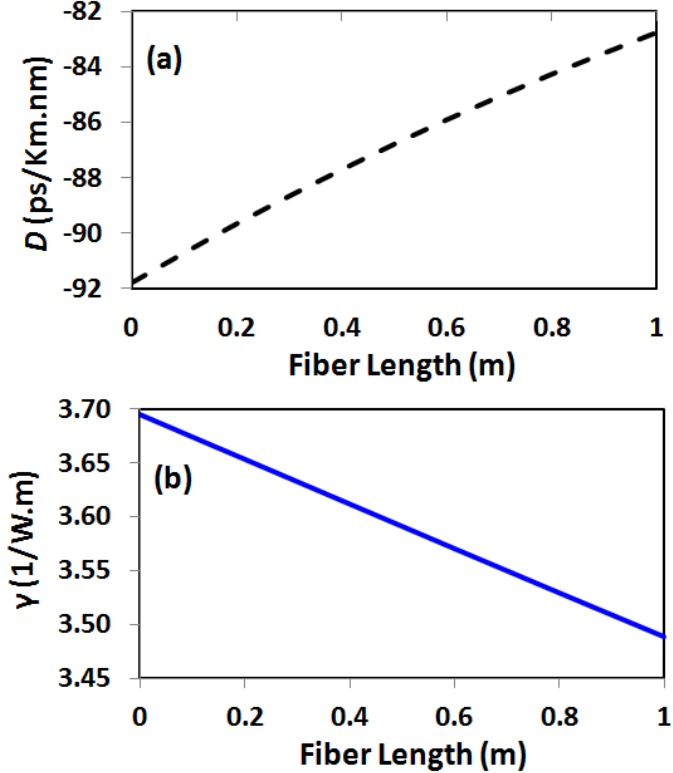

Fig. 2. Variation of (a) Dispersion (*D*) and (b) nonlinear parameter (*γ*) along fiber length. (color online)

### B. Fiber Length Optimization

In order to check the quality of generated PP from input Gaussian pulse, we compute the evolution of misfit parameter (*M*) between the output intensity, $|U(T)|^2$ and best fit parabolic intensity, $|P(T)|^2$ as described in [10] through

$$M^2 = \frac{\int \left[|U|^2 - |P|^2\right]^2 dT}{\int |U|^4 dT} \qquad (4)$$

Optimum fiber length is identified where this *M* become minimum. The variation of this *M* along the fiber length is shown in Fig. 3. *M* reaches its minimum value (0.0216) after ~ 19 cm length of propagation.

### C. Pulse Evolution

Tm$^{3+}$-doped pulsed fiber laser operating at 2.04 μm with a peak power ($P_{peak}$) of 60 W and FWHM of 3.5 ps is assumed as the input to the designed MOF. The evolution of this input Gaussian pulse along the fiber length is shown in Fig. 4. Where the blue color represents the input Gaussian pulse and the red color represents the parabolic output. The output pulse FWHM becomes ~ 4.98 ps with peak normalized power of ~ 0.775 (46.48 W).

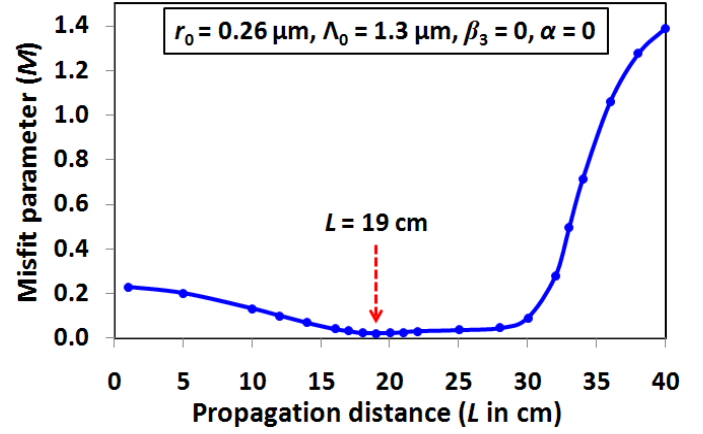

Fig. 3. Variation of misfit parameter (*M*) along the length of the linearly tapered MOF. *M* reaches its minimum value at a *L* ~ 19 cm. (color online)

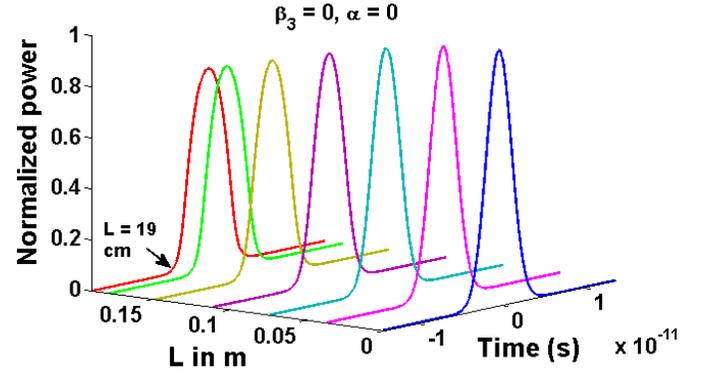

Fig. 4. Evolution of an input Gaussian pulse (blue color) to a PP (red color) in the loss less designed tapered fiber of 19 cm length. (color online)

Reshaping of a Gaussian pulse to the desired PP is also evident from the plot of output power profile (cf. Fig. 5).

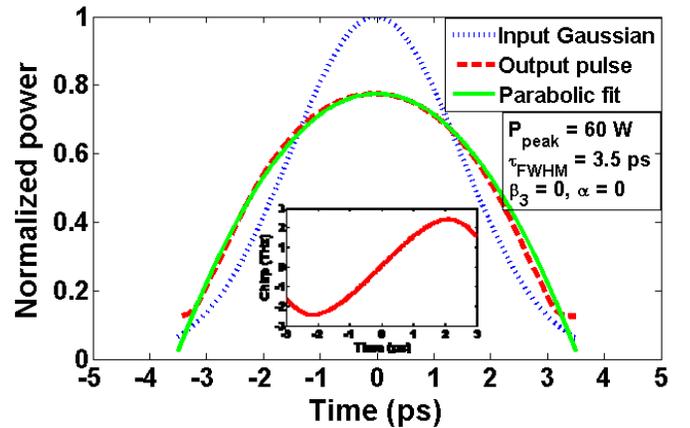

Fig. 5. Output pulse shape (red dashed) from the 19 cm long designed tapered MOF, fitted by an ideal parabolic shape (green solid). Inset showing the linear chirp across its temporal profile. (color online)

The blue dotted curve, red dashed curve and green solid curve correspond to Gaussian input, parabolic output and a best fit ideal PP, respectively. Misfit parameter is 0.0216 which is quite small. The inset of Fig. 5 shows the unique characteristic of linear chirp acquired over the duration of the generated PP.

## IV. RESULTS INCLUDING TOD AND LOSS

Until now we have assumed a loss-less fiber and neglected any TOD effect along its length. However, loss as well as TOD could degrade the quality of the generated PP. In the following, we describe results of our analysis of PP generation by including these effects.

For the designed fiber, the variation of TOD ($\beta_3$) along tapered MOF length is shown in Fig. 6, where the value of $\beta_3$ is almost linearly varied from 0.049 ps$^3$/km to 0.9 ps$^3$/km over 19 cm length of the tapered MOF. The output normalized power profile *corresponding to* $\beta_3 = 0$ (blue solid) and $\beta_3 \neq 0$ (red dotted) are shown in Fig. 7. One inset (inset (a)) of Fig. 7 shows the linear chirp across its temporal profile for both the cases. On the other hand, as the two power profile almost overlap with each other, we have plotted the difference between these two powers ($\Delta P$) as a second inset (inset (b)) (green dashed), where $\Delta P$ remains below 0.04% for the entire pulse duration. As this TOD effect is negligibly small, higher order dispersion effects would be smaller and hence can be neglected. Thus it is safe to conclude that our designed tapered fiber will generate distortion free parabolic pulses.

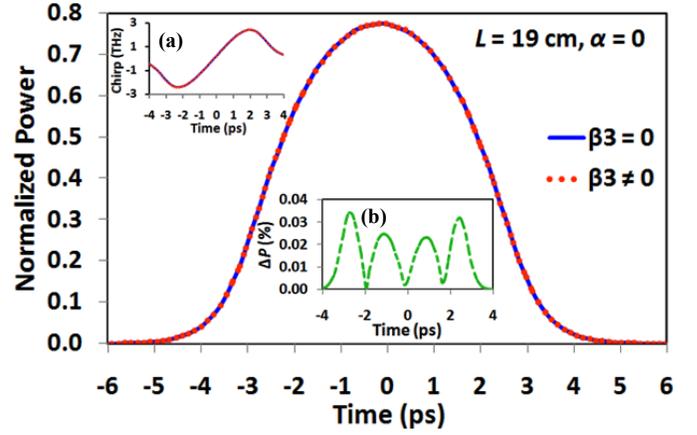

Fig. 7. Output pulse shape including $\beta_3$ effect (red dotted) and neglecting $\beta_3$ effect (blue solid). Insets show (a) linear chirp for both the cases ($\beta_3 = 0$ and $\beta_3 \neq 0$) and (b) difference between two output powers ($\Delta P$) in % (green dashed). (color online)

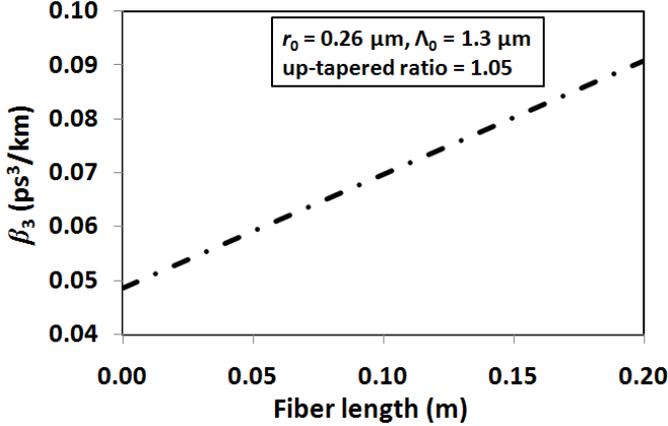

Fig. 6. Variation of $\beta_3$ along tapered fiber length.

In the next step we studied the effect of loss on the generated PP in the presence of TOD. We considered both the material loss (~ 0.4 dB/m at 2 μm) and confinement loss ($\alpha_c$). For our designed MOF, the $\alpha_c$ is ~ 1 dB/m. So, the total loss ($\alpha$) is ~ 1.4 dB/m. Output normalized power profile for $\alpha = 0$ (blue solid) and $\alpha \neq 0$ (red dashed) is shown in Fig. 8. Inset shows the variation in chirp across the temporal profile. For both the cases, chirp is linear, which confirms PP output even after inclusion of loss. Incidentally this loss reduces the maximum output power by only 4.4 %. Loss management is an important issue to address as loss of the fabricated fiber will be typically higher than that of the designed one due to potentially unavoidable local deformations that may occur during fabrication of the taper.

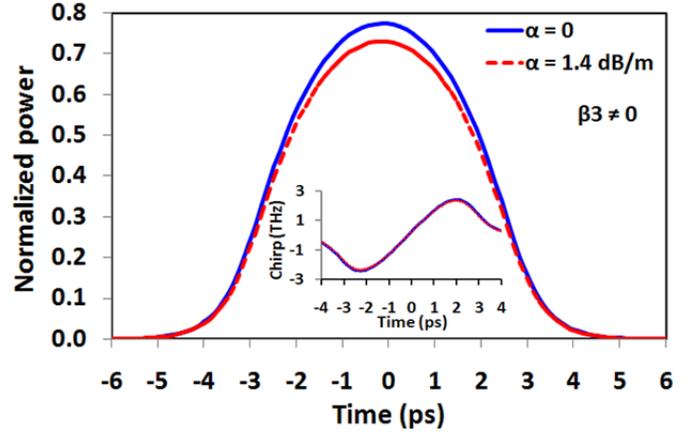

Fig. 8. Output pulse shape after including loss (red dashed) and without loss (blue solid). For both the plots, TOD is included. (color online)

## V. EFFECTS OF INPUT PULSE PARAMETERS

Further we studied the effects of input pulse parameters like peak power, pulse width, pulse energy etc. on output pulse shape after including TOD and loss.

### A. Effect of Input Peak Power ($P_{peak}$)

By varying peak power ($P_{peak}$) of the input Gaussian pulse from 10 W to 100 W, we have studied the variation of optimum fiber length ($L_{opt}$) corresponding to minimum $M^2$ value, output pulse width (output $\tau_{FWHM}$) and maximum normalized output power for a fixed input $\tau_{FWHM}$ of 3.5 ps. These variations are shown in Figs. 9 (a) – (c), respectively. $L_{opt}$ gradually decreases with increasing the input $P_{peak}$ (cf. Fig. 9 (a)), which implies that, with increasing peak power, output pulse gets its parabolic shape at shorter propagation distance. Figures 9 (b) and (c) show small variation in the output $\tau_{FWHM}$ and maximum normalized power for $L = L_{opt}$, however, at our designed length (19 cm), output $\tau_{FWHM}$ gradually increases and maximum power gradually decreases.

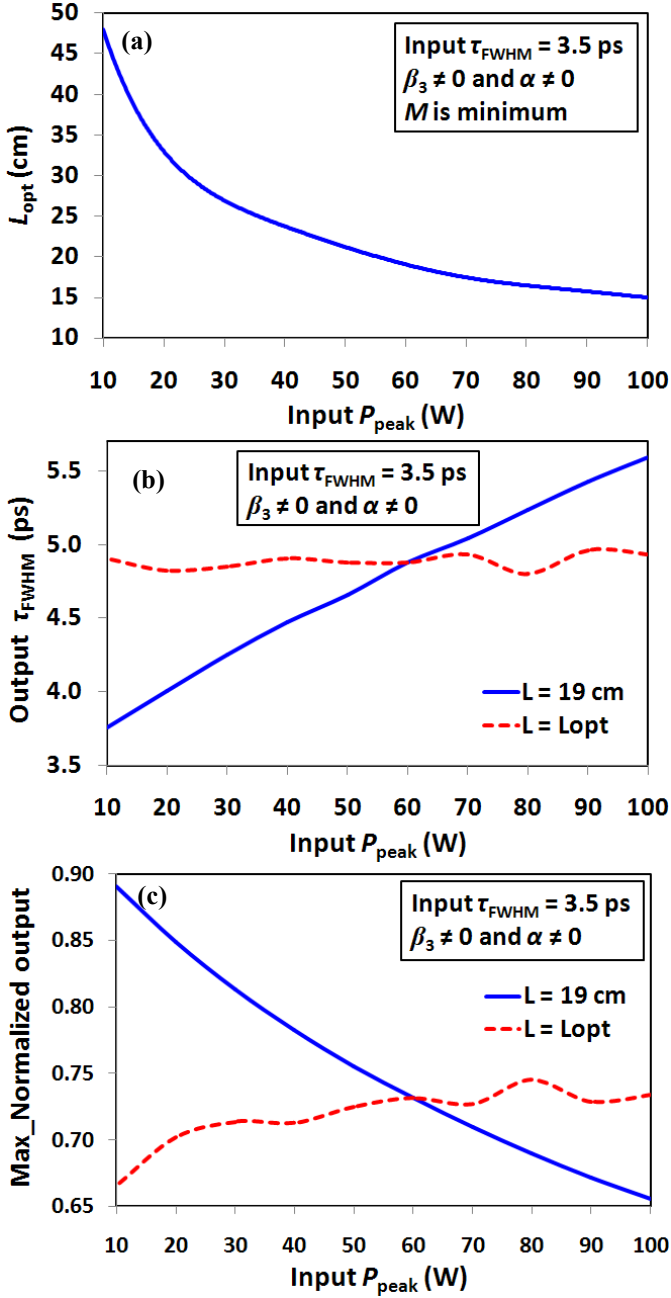

Fig. 9. Variations of (a) optimum fiber length ($L_{opt}$); (b) Output pulse width ($\tau_{FWHM}$); and (c) Maximum normalized output power with the input peak power ($P_{peak}$). For (b) and (c), the blue solid line and red dashed line correspond to the value for the designed tapered fiber length ($L = 19$ cm) and at $L_{opt}$ for that particular input $P_{peak}$, respectively. (color online)

The shape of output pulse for $L = L_{opt}$ is shown in Fig. 10 for three different $P_{peak}$ (40 W, 60 W and 80 W) with input $\tau_{FWHM}$ of 3.5 ps. Their chirp characteristics are also shown in the inset of Fig. 10.

### B. Effect of Input Pulse Width (Input $\tau_{FWHM}$)

Unlike the effect of input peak power, $L_{opt}$ gradually increases with increase in input $\tau_{FWHM}$ (cf. Fig. 11 (a)) for a fixed input $P_{peak}$ of 60 W. In Fig. 11 (b), variation of the ratio of output and input pulse widths ($\tau_{Norm}$) is shown for $L = L_{opt}$ (red dashed) and for $L = 19$ cm (blue solid). Though $\tau_{Norm}$ decreases with input pulse width for our designed length, it remains almost constant for $L = L_{opt}$. On the other hand, the maximum normalized output power increases rapidly for $L = 19$ cm (blue solid line in Fig. 11 (c)) where as for the optimum length it decreases very slowly with increasing input $\tau_{FWHM}$ (red dashed in Fig. 11 (c)).

Only the output pulse shape for $L = L_{opt}$ is shown here in Fig. 12 for three different values of input $\tau_{FWHM}$ (2.5 ps, 3.5 ps, and 4.5 ps) for fixed input $P_{peak}$ of 60 W. The inset shows their chirp characteristics.

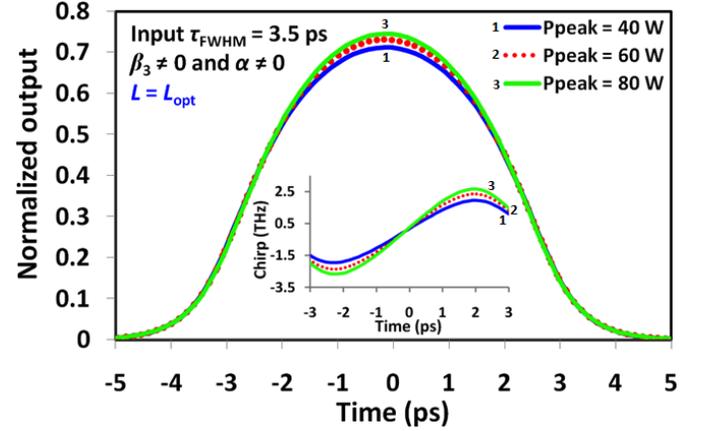

Fig. 10. Output pulse shape for $P_{peak}$ = 40 W, 60 W and 80 W at their optimum lengths for $\tau_{FWHM}$ of 3.5 ps. Inset show variations in their chirp with time. (color online)

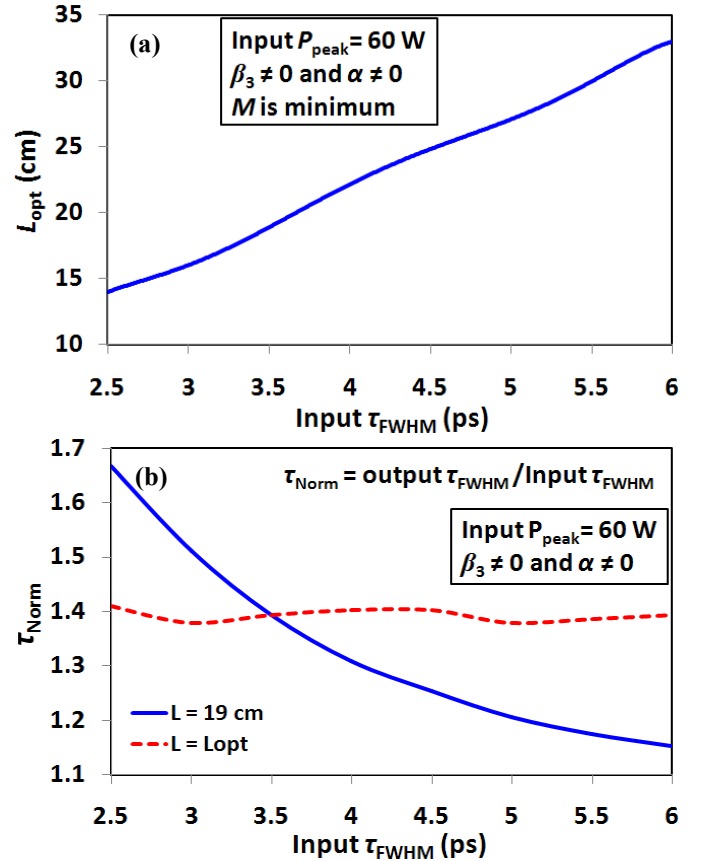

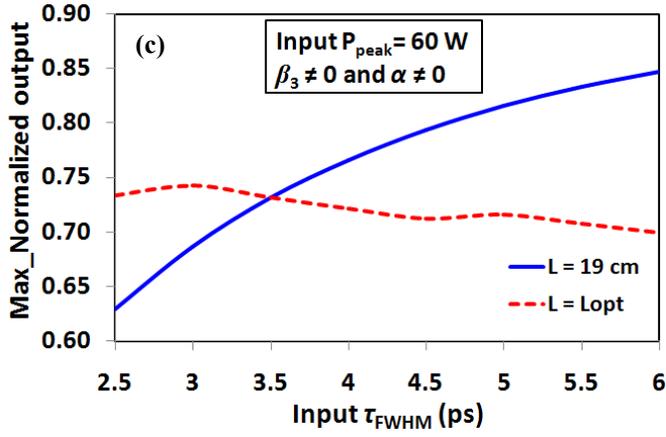

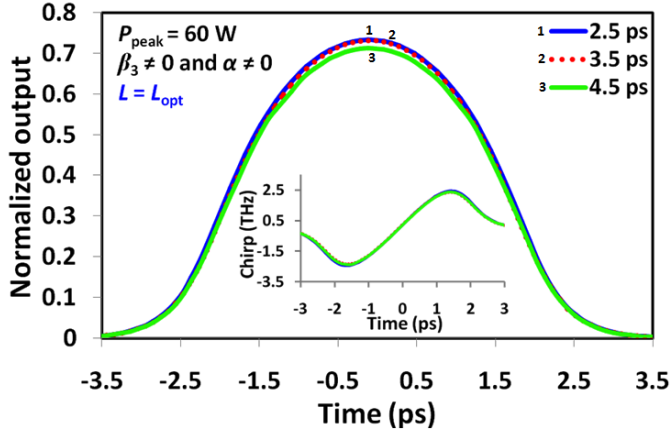

Fig. 11. Variation of (a) optimum fiber length ($L_{opt}$); (b) Normalized output pulse width ($\tau_{Norm}$); and (c) Maximum normalized output power with input $\tau_{FWHM}$. For (b) and (c), the blue solid line and the red dashed line correspond to the value for our designed fiber length ($L$ = 19 cm) and at $L_{opt}$ for that particular input $\tau_{FWHM}$, respectively. (color online)

Fig. 12. Output pulse shape for input $\tau_{FWHM}$ = 2.5 ps, 3.5 ps and 4.5 ps at the end of corresponding optimum lengths for a $P_{peak}$ of 60 W. Inset shows variations in their chirp with time. (color online)

### C. Results for Fixed Input Pulse Energy (Input $E_{pulse}$)

The peak power ($P_{peak}$) of a Gaussian pulse is ≈ 0.94 times the pulse energy ($E_{pulse}$) divided by the FWHM pulse duration ($\tau_{FWHM}$) [31]. Previously we have studied the effect on output PP by varying either $P_{peak}$ or $\tau_{FWHM}$ of the input pulse. Essentially we varied the pulse energy in both the cases. Then we studied the effect of a fixed $E_{pulse}$ on output for different sets of input $P_{peak}$ and $\tau_{FWHM}$. For $E_{pulse}$ = 0.2234 nJ that corresponds to a $P_{peak}$ of 60 W and $\tau_{FWHM}$ of 3.5 ps, the variation of optimum length is studied and shown in Fig. 13 (a), where $L_{opt}$ rapidly decreases from almost 100 cm to 10 cm for a change in $P_{peak}$ from 20 W to 90 W. The variation of $\tau_{Norm}$ and maximum normalized output power for fixed $E_{pulse}$ (0.2234 nJ) are shown in Figs. 13 (b) and (c), respectively. Fig. 13 (b) implies that for our designed fiber, normalized output pulse width become maximum (almost 1.7 x input $\tau_{FWHM}$) for $P_{peak}$ of ~ 80 W. So automatically, the output peak power becomes lowest for this case, hence we get a dip around an input $P_{peak}$ of 80 W (cf. Fig. 13 (c)).

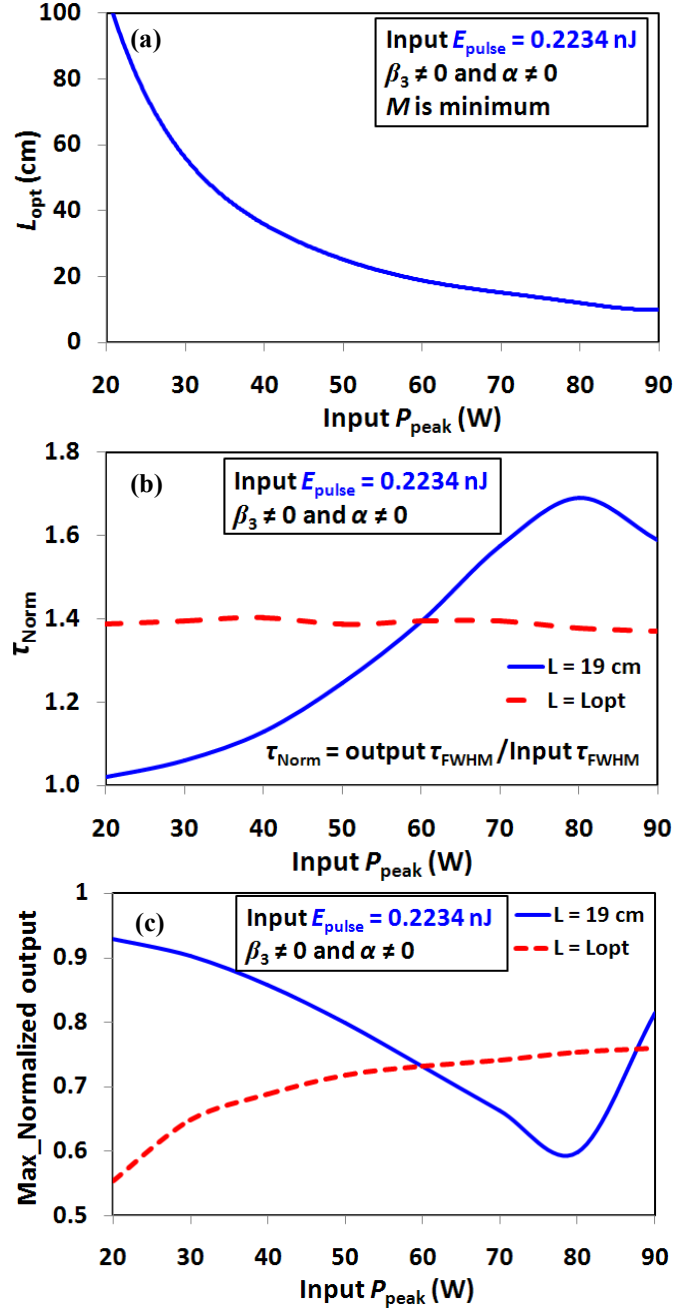

Fig. 13. Variation of (a) Optimum fiber length ($L_{opt}$); (b) Normalized output pulse width ($\tau_{Norm}$); and (c) Maximum normalized output power with the input peak power ($P_{peak}$) for fixed pulse energy of 0.2234 nJ. For (b) and (c), the blue solid line and red dashed line correspond to the value at our designed fiber length ($L$ = 19 cm) and at $L_{opt}$ corresponding to that particular set of input $P_{peak}$ and $\tau_{FWHM}$, respectively. (color online)

We have also plotted output pulse shapes for three different sets of input $P_{peak}$ and $\tau_{FWHM}$ (40 W and 5.25 ps, 60 W and 3.5 ps, 80 W and 2.625 ps) (cf. Fig. 14) at the end of their corresponding optimum lengths; inset in Fig. 14 shows variations in their chirp with time.

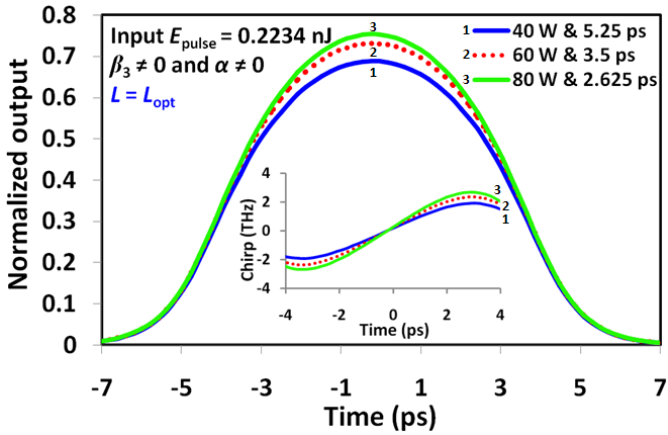

Fig. 14. Output pulse shape for input $P_{peak}$ and $\tau_{FWHM}$ of 40 W and 5.25 ps (blue solid), 60 W and 3.5 ps (red dotted), 80 W and 2.625 ps (green solid) with fixed $E_{pulse}$ of 0.2234 nJ. Inset shows variations in their chirp with time. (color online)

## VI. Conclusion

Through a detailed numerical study we present a realistic design of chalcogenide glass based up-tapered MOF structure to generate high power, breaking free parabolic pulses (PPs) within a very short length of our designed MOF for a Gaussian input. MOF was chosen to consist of $As_2S_3$ glass with four rings of hexagonally arranged holey cladding. Air hole separation ($\Lambda$) and their diameter ($d$) were optimized to get proper dispersion profile, single mode operation, and lowest loss at the operating wavelength. Up-tapered structure with taper ratio of 1.05 acts as a passive dispersion decreasing fiber suitable for PP generation free from distortion and noise. The effect of TOD and loss were found to be negligible on the generated pulse quality. In summary, for a Gaussian input pulse of 60 W peak power with FWHM of 3.5 ps, we have shown that after propagation through only 19 cm of the designed tapered MOF one can realize a parabolic output pulse with normalized peak power of 0.78 (46.48 W), and of pulse FWHM of 4.98 ps at a wavelength of 2.04 μm. Further, the effects of input pulse parameters on the generated PP were also studied to check its operating regime.

PPs should be very useful for pulse shaping, pulse synthesis, high power handling, supercontinuum source generation, etc. at mid-IR wavelength region, and also for medical diagnostics.

This study can be further extended by investigating roles of different input pulse shapes and taper profiles, and choosing appropriately suitable fiber material(s) for other operating wavelength regimes.

*Some salient features of these results were recently reported by us at the international conference ICMAT 2013 from June 30 to July 5, 2013 at Suntec, Singapore.*